\documentclass[fleqn,usenatbib]{mnras}
\usepackage{mathptmx}
\usepackage{txfonts}
\usepackage[T1]{fontenc}
\usepackage{ae,aecompl}
\usepackage{graphicx}

\usepackage{amssymb}

\usepackage{natbib}
\usepackage{color}

\title[Young stars in the LMC periphery]{Young stars in the periphery of the Large Magellanic Cloud}

\author[C. Moni Bidin et al.]{
C. Moni Bidin,$^{1}$\thanks{E-mail: cmoni@ucn.cl}
D.~I. Casetti-Dinescu,$^{2}$
T.~M. Girard,$^{3}$
L. Zhang,$^{4,5}$
\newauthor{
R.~A. M\'endez,$^{6}$
K. Vieira,$^{7}$
V.~I. Korchagin,$^{8}$
W.~F. van Altena$^{9}$}
\\
$^{1}$Instituto de Astronom\'ia, Universidad Cat\'olica del Norte, Av. Angamos 0610, Antofagasta, Chile\\
$^{2}$Department of Physics, Southern Connecticut State University, 501 Crescent Street, New Haven, CT 06515, USA\\
$^{3}$14 Dunn Rd, Hamden, CT 06518, USA\\
$^{4}$CAS South America Center for Astronomy, Camino El observatorio 1515, Las Condes, Santiago, Chile\\
$^{5}$Key Lab of Optical Astronomy, National Astronomical Observatories, CAS, 20A Datun Road, Chaoyang District, 100012 Beijing, China\\
$^{6}$Departamento de Astronom\'ia, Universidad de Chile, Casilla 36-D, Santiago, Chile\\
$^{7}$Centro de Investigaciones de Astronom\'ia, Apartado Postal 264, M\'erida 5101-A, Venezuela\\
$^{8}$Institute of Physics, Southern Federal University, Stachki Street 124, 344090, Rostov-on-Don, Russia\\
$^{9}$Astronomy Department, Yale University, P.O. Box 208101, New Haven, CT 06520-8101, USA
}

\date{Accepted XXX. Received YYY; in original form ZZZ}
\pubyear{2016}
\begin{document}
\label{firstpage}
\pagerange{\pageref{firstpage}--\pageref{lastpage}}
\maketitle

\begin{abstract}
Despite their close proximity, the complex interplay between the two Magellanic Clouds, the Milky Way, and the resulting tidal features, is
still poorly understood. Recent studies have shown that the Large Magellanic Cloud (LMC) has a very extended disk strikingly perturbed in its
outskirts. We search for recent star formation in the far outskirts of the LMC, out to $\sim30\degr$ from its center. We have collected
intermediate-resolution spectra of thirty-one young star candidates in the periphery of the LMC and measured their radial velocity, stellar
parameters, distance and age. Our measurements confirm membership to the LMC of six targets, for which the radial velocity and distance values
match well those of the Cloud. These objects are all young (10--50~Myr), main-sequence stars projected between 7$\degr$ and 13$\degr$ from the
center of the parent galaxy. We compare the velocities of our stars with those of a disk model, and find that our stars have low to moderate
velocity differences with the disk model predictions, indicating that they were formed {\it in situ}. Our study demonstrates that recent star
formation occurred in the far periphery of the LMC, where thus far only old objects were known. The spatial configuration of these
newly-formed stars appears ring-like with a radius of 12~kpc, and a displacement of 2.6~kpc from the LMC's center. This structure, if real,
would be suggestive of a star-formation episode triggered by an off-center collision between the Small Magellanic Cloud and the LMC's disk.
\end{abstract}



\section{Introduction}
\label{s_intro}

The Large and Small Magellanic Clouds (LMC and SMC, respectively) form a binary pair of gas-rich, star-forming galaxies that are satellites
of the Milky Way, located within $\sim50$~kpc of their host. This is a configuration that is rarely found around Milky-Way type galaxies
\citep{James11,Liu11}. Interactions between the Clouds themselves and with the Galaxy have generated a complex ensemble of gaseous
structures such as: the extended Magellanic Stream \citep{Mathewson74}, encompassing $\sim200\degr$ across the sky along the Clouds' orbit
\citep{Nidever10}; the Bridge between the Clouds \citep{Kerr57,Hindman63}; and the Leading Arm (LA), a fragmented, wide feature leading
the orbit of the Clouds \citep[][see their Fig.~5]{Gardiner96}. Observational data on the gas in this system have accumulated over the past
decades and it is now realized that the Magellanic Stream, Bridge and LA bring to the halo of the Milky Way a mass of gas on the order of
$2\times10^9$~M$_{\odot}$, or over twice as much gas as the Clouds themselves contain \citep[see review by][]{Donghia15}. Are there any stars
in these gaseous features? Tidal models predict the existence of stars, however observations have not been decisive primarily due to the lack
of deep, wide and contiguous photometric surveys. While young, recently-formed stars have been found in the Bridge
\citep{Irwin90,Dem98,Harris07,Dinescu12,Skowron14} and, more recently, in the LA \citep{Dinescu14}, a substantial, intermediate-age population
is still elusive, with some hints of its presence in the Bridge \citep{Bagheri13,Saha15,Noel15}. This status is however improving, primarily
due to the Dark Energy Survey (DES), which has recently revealed a large, complex stellar extent to the LMC \citep{mackey16,Belokurov16} as
well as new dwarf galaxies, some of which may be associated with the Magellanic system \citep[see review by][]{Donghia15}.

In this contribution, we focus on the outskirts of the LMC. Earlier studies based on near-IR surveys agree that the LMC disk extends to some
9~kpc in radius (or about $8\degr$ on the sky), and is made of an intermediate to old stellar population. However, \citet{Munoz06} report
the finding of a handful of giant stars, likely members of the LMC, in the field of Carina dwarf galaxy some $20\degr$ away from the LMC's
center. This finding was later confirmed by \citet{McMonigal14}. \citet{Saha10} present a photometric study in a series of fields that
probes the outer regions of the LMC. In particular, the northward fields from a distance of $7\degr$ to $20\degr$ from the LMC's center show
an exponential decline in the number of stars; this indicates that the intermediate-age/old disk extends out to $16\degr$ from the LMC's
center \citep[see][and references therein]{Saha10}. Finally, the recent study by \citet{mackey16} finds in the same northern region of the
LMC a 10-kpc tidal arm at some $16\degr$ from the LMC's center. Clearly the disk of the LMC is very extended and perturbed in its outer parts.

Given that the disk extends so far out, and that it shows evidence of tidal interaction in its outskirts, we seek to know just how far from
LMC's center star formation can occur. This will help us better understand star formation in an environment that was subject to a recent
($\sim200$~Myr) interaction with the SMC, i.e., a turbulent and tidally stressed environment. To this end, we search for young, newly-formed
stars that are LMC members.
\defcitealias{Dinescu12}{CD12}
\citet[][hereafter CD12]{Dinescu12} listed 567 OB star candidates in a wide area including the periphery of the Clouds, the Bridge, the
Leading Arm, and part of the Magellanic Stream. These stars were selected from cuts in magnitude, colors, and proper motions, after merging
the UV photometry from the Galaxy Evolution Explorer \citep[GALEX;][]{Bianchi11}, the infrared data of the Two Micron All-Sky Survey
\citep[2MASS;][]{Skrutskie06}, the optical photometry of the AAVSO Photometric All-Sky Survey \citep[APASS;][]{Henden11}, and proper motions
and photometry from the Southern Proper Motion program \citep[SPM4;][]{Girard11}. The selection criteria were adjusted to isolate distant,
hot, young main-sequence (MS) stars. However, hot subdwarf stars \citep[sdB's,][]{Heber86}, and white dwarfs (WDs) should inevitably be
present in the sample, because they are photometrically indistinguishable from hot MS stars. These objects are intrinsically fainter than MS
stars by more than five magnitudes, hence any such star in the sample will be a foreground Galactic object. Galactic runaway stars could
also contaminate the sample to a lesser extent.

We started an extensive spectroscopic campaign to investigate the \citetalias{Dinescu12} young star candidates in more detail \citep{Moni13}.
In \citet{Dinescu14}, we identified some 19 young OB stars in the direction of the LA, where no stellar component was previously known; five
of these have kinemactics consistent with LA membership and are located at the edge of the Galactic disk. This discovery suggests that the
interaction between the Clouds and our Galaxy is strong enough to trigger recent star formation in certain regions of this gaseous structure.
However, our analysis also revealed that a relevant fraction (55\%) of the candidate sample consists of foreground, more compact Galactic
objects. In this paper, we focus on the young star candidates found in the periphery of the LMC, i.e. out to a projected radius of about
33$\degr$ from the LMC center, while excluding the inner 6$\degr$.


\section{Observations and measurements}
\label{s_data}

\subsection{Observations and data reduction}
\label{ss_obs}

We selected thirty-one young star candidates with $V<16$ between 6$\degr$ and 33$\degr$ from the center of the LMC, from the catalog of
\citet{Dinescu12}. The IDs, coordinates, and $V$ magnitudes of target stars, drawn from the SPM4 catalog \citep{Girard11}, are given in
the first four columns of Table~\ref{t_data}. Their intermediate-resolution spectra (R$\approx$3500) were collected during three
half-nights of observations between 2015 January~28 and February~22 at Las Campanas Observatory, with the IMACS spectrograph at the
focus of the Baade 6.5m telescope. The instrument was used at f/4 in longslit mode, and the 0$\farcs$75-wide slit was employed. The
1200--17.5 grating was tilted by an angle of 16$\fdg$8 to cover the spectral range 3660--5250~\AA\ on the CCD. Exposure times varied
between 150s and 1500s, according to target magnitude and sky conditions, to yield spectra with signal-to-noise ratio S/N$>50$. A
comparison lamp for wavelength calibration was acquired after each exposure. A twilight solar spectrum was collected at the beginning
of each night, and the radial-velocity standards LTT2415 \citep{Latham02} and GJ273 \citep{Chubak12} were also observed.

\begin{table*}
\centering
\caption{Data and measurements of target stars.}
\label{t_data}
\begin{tabular}{ccccrlclll}
\hline
\hline
SPM ID & RA & Dec & $V$ & RV & T$_\mathrm{eff}$ & $\log{(g)}$ & $\log{(\frac{N(He)}{N(H)})}$ & $V\sin(i)$ & note \\
 & h:mm:ss.s & $\degr:\arcmin:\arcsec$ & & km~s$^{-1}$ & K & dex & dex & km~s$^{-1}$ & \\
\hline
0030025059 & 3:32:14.6 & $-$83:44:13 & 14.08 & 183$\pm$6 \ \   & 17000$\pm$250  & 3.88$\pm$0.06 & $-$1.00          & $<50$     & MS \\
0870012815 & 3:42:24.1 & $-$64:03:21 & 14.82 & 19$\pm$7 \ \    & 18800$\pm$350  & 4.69$\pm$0.06 & $-$2.20$\pm$0.09 & $\sim50$  & sdB/WD \\
0140123632 & 3:47:26.8 & $-$78:41:54 & 15.90 &  413$\pm$11     & 15900$\pm$500  & 3.81$\pm$0.09 & $-$2.15$\pm$0.15 & $\sim50$  & post-HB? \\
0570014281 & 3:50:37.7 & $-$69:20:57 & 14.70 & $-$24$\pm$15    & 42600$\pm$200  & 6.19$\pm$0.09 & $+$0.15$\pm$0.09 & $\sim50$  & sdB/WD \\
0320149401 & 4:13:50.7 & $-$72:18:52 & 15.95 &    26$\pm$10    & 25300$\pm$500  & 5.55$\pm$0.06 & $-$2.83$\pm$0.12 & $<50$     & sdB/WD \\
0320022407 & 4:15:21.2 & $-$72:51:41 & 13.93 &  $-$4$\pm$7 \ \ & 24100$\pm$500  & 5.18$\pm$0.06 & $-$2.56$\pm$0.12 & $<50$     & sdB/WD \\
1660012832 & 4:15:30.2 & $-$54:21:59 & 14.91 &  $-$7$\pm$21    & 50600$\pm$700  & 6.50$\pm$0.09 & $+$0.34$\pm$0.27 & $<50$     & sdB/WD \\
0880011003 & 4:19:30.8 & $-$65:22:13 & 13.98 &    12$\pm$6 \ \ & 25300$\pm$500  & 5.02$\pm$0.06 & $-$2.08$\pm$0.09 & $<50$     & sdB/WD \\
0320115607 & 4:19:55.5 & $-$73:51:45 & 16.04 &   164$\pm$9 \ \ & 24500$\pm$400  & 4.03$\pm$0.06 & $-$1.03$\pm$0.09 & $\sim100$ & MS \\
0320018938 & 4:19:58.5 & $-$73:52:26 & 14.95 &   193$\pm$8 \ \ & 22300$\pm$400  & 3.78$\pm$0.09 & $-$1.02$\pm$0.09 & $\sim100$ & MS \\
0040202213 & 4:20:32.1 & $-$82:06:35 & 15.89 & $-$27$\pm$21    & 47500$\pm$800  & 6.14$\pm$0.15 & $+$0.78$\pm$0.18 & $<50$     & sdB/WD \\
1250013582 & 4:26:29.2 & $-$59:06:10 & 14.29 & $-$17$\pm$6 \ \ & 26000$\pm$500  & 5.41$\pm$0.06 & $-$2.84$\pm$0.09 & $<50$     & sdB/WD \\
1250015381 & 4:36:09.8 & $-$58:37:07 & 14.90 &   212$\pm$7 \ \ & 21100$\pm$400  & 3.50$\pm$0.09 & $-$0.95$\pm$0.09 & $\sim100$ & MS \\
2740050272 & 4:42:20.2 & $-$45:52:29 & 15.73 &    50$\pm$12    & 59000$\pm$3000 & 6.01$\pm$0.12 & $-$2.22$\pm$0.15 & $<50$     & sdB/WD \\
1270012151 & 5:16:10.6 & $-$60:57:37 & 14.81 &     2$\pm$6 \ \ & 25700$\pm$400  & 5.40$\pm$0.06 & $-$2.66$\pm$0.09 & $<50$     & sdB/WD \\
1680007950 & 5:29:01.2 & $-$56:33:10 & 15.06 &    43$\pm$7 \ \ & 27400$\pm$400  & 5.50$\pm$0.06 & $-$2.53$\pm$0.09 & $<50$     & sdB/WD \\
0340108821 & 5:57:02.2 & $-$75:40:23 & 16.04 & $-$52$\pm$11    & - & - & - & - & sdB/WD \\
1280025746 & 5:58:49.6 & $-$58:56:27 & 15.25 &   303$\pm$5 \ \ & 21700$\pm$350  & 3.65$\pm$0.06 & $-$1.00          & $<50$     & MS \\
1280143250 & 6:00:00.6 & $-$59:01:03 & 16.01 & $-$34$\pm$9 \ \ & 42200$\pm$300  & 5.62$\pm$0.15 & $+$0.58$\pm$0.09 & $<50$     & sdB/WD \\
1280031295 & 6:00:22.1 & $-$57:54:20 & 15.54 &   368$\pm$8 \ \ & 16000$\pm$250  & 3.33$\pm$0.06 & $-$1.00          & $\sim50$  & MS \\
0910042645 & 6:05:34.6 & $-$62:55:06 & 15.66 & $-$12$\pm$7 \ \ & 27800$\pm$400  & 5.48$\pm$0.06 & $-$2.89$\pm$0.12 & $<50$     & sdB/WD \\
3380011095 & 6:15:47.1 & $-$41:32:06 & 14.65 &   424$\pm$6 \ \ & 16100$\pm$200  & 4.05$\pm$0.06 & $-$1.00          & $<50$     & MS \\
1280035162 & 6:20:38.5 & $-$57:05:38 & 14.72 & $-$11$\pm$5 \ \ & 34100$\pm$500  & 5.60$\pm$0.06 & $-$0.89$\pm$0.06 & $<50$     & sdB/WD \\
0350036197 & 6:46:44.1 & $-$72:10:37 & 13.64 &    23$\pm$5 \ \ & 29400$\pm$900  & 5.14$\pm$0.12 & $-$1.16$\pm$0.09 & $<50$     & sdB/WD \\
0600320649 & 6:48:41.5 & $-$68:30:36 & 16.14 &  $-$4$\pm$16    & 42400$\pm$300  & 5.66$\pm$0.15 & $+$0.55$\pm$0.09 & $<50$     & sdB/WD \\
0600047219 & 6:53:13.1 & $-$66:59:51 & 14.64 &    55$\pm$9 \ \ & 37500$\pm$900  & 5.38$\pm$0.12 & $-$3.16$\pm$0.18 & $<50$     & sdB/WD \\
0920126094 & 6:59:58.0 & $-$66:18:28 & 16.38 &   284$\pm$7 \ \ & 19100$\pm$400  & 3.78$\pm$0.06 & $-$1.00          & $\sim100$ & MS \\
0350232289 & 7:04:25.2 & $-$72:38:43 & 15.89 &   280$\pm$8 \ \ & 22400$\pm$600  & 3.69$\pm$0.09 & $-$1.00          & $\sim200$ & MS \\
0350031043 & 7:10:55.1 & $-$73:27:21 & 15.18 & $-$35$\pm$7 \ \ & 27100$\pm$600  & 5.39$\pm$0.09 & $-$2.64$\pm$0.12 & $<50$     & sdB/WD \\
1310017952 & 7:48:37.5 & $-$61:50:03 & 14.91 &     8$\pm$6 \ \ & 27000$\pm$500  & 5.45$\pm$0.06 & $-$2.18$\pm$0.09 & $<50$     & sdB/WD \\
0950048906 & 8:39:21.8 & $-$64:49:35 & 15.55 &    27$\pm$8 \ \ & 32900$\pm$400  & 5.46$\pm$0.06 & $-$3.33$\pm$0.12 & $<50$     & sdB/WD \\
\hline
\end{tabular}
\end{table*}

\begin{figure}
\includegraphics[width=\columnwidth,angle=-90]{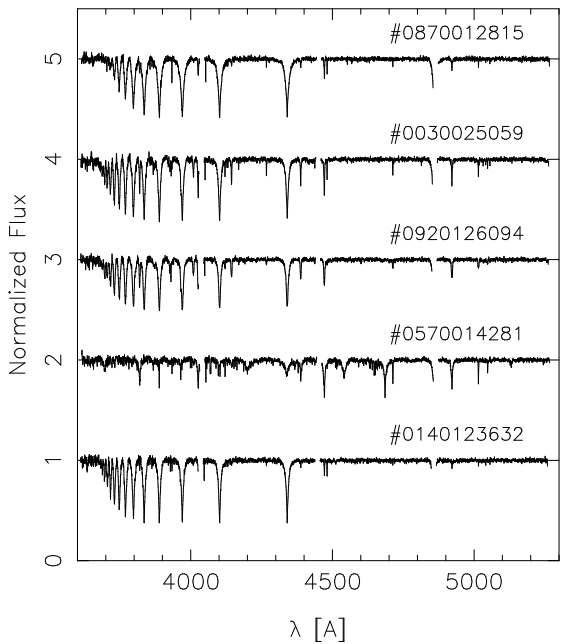}
\caption{Examples of reduced spectra of our sample stars. The spectra are shifted vertically to avoid overlap. The SPM ID of each target is
also indicated. {From top to bottom, we show the spectrum of a helium-poor sdB, a MS star at the same temperature, a fast-rotating MS object,
a hot WD, and the uncertain SPM4 object 0140123632 discussed in Sect.~\ref{ss_sdb}.}}
\label{f_spec}
\end{figure}

The spectra were split on four chips with short gaps between them. Each spectral section was independently reduced, extracted with an
optimum algorithm \citep{Horne86} and normalized, with standard IRAF\footnote{IRAF is distributed by the National Optical Astronomy
Observatories, which are operated by the Association of Universities for Research in Astronomy, Inc., under cooperative agreement with the
National Science Foundation.} routines \citep[see, e.g.,][]{delaFuentes15}. Before merging them, the absolute offsets of the four
independent wavelength calibration solutions were measured on the standard stars, and corrected. Some reduced spectra are shown in
Fig.~\ref{f_spec} as an example.

\subsection{Radial velocities}
\label{ss_rv}

We measured the radial velocity (RV) of target stars using IRAF task {\it fxcor}, cross-correlating \citep{Simkin74,Tonry79} their spectra
with synthetic templates drawn from the library of \citet{Munari05}. Previous investigations have shown that the RV measurements are not
affected by the exact choice of the template, because a partial mismatch between object and template spectral types only enhances the formal
uncertainties, without shifting the peak of the cross-correlation function \citep{Morse91,Moni11}. All RVs were corrected to heliocentric
values. The results are given in the fifth column of Table~\ref{t_data}. The errors were computed as the quadratic sum of the most relevant
sources. The wavelength calibration introduced a negligible uncertainty, smaller than 1~km~s$^{-1}$ for all stars. The cross-correlation error
was in the range 2--8~km~s$^{-1}$ (but up to 15--20~km~s$^{-1}$ in four cases), and thus it usually dominated the error budget. Some
systematic offsets are expected, caused by an imperfectly-centered position of the star in the slit \citep[see the discussion in][]{Moni06}.
These are hard to estimate and remove, in absence of deep telluric lines in the observed spectral range. Instead, they were estimated from
the rms of the residual RV offsets of the standard stars, and thus approximated as a fixed 4~km~s$^{-1}$ for all targets. The spectra were
eventually shifted to laboratory wavelength for the measurement of the stellar parameters.

\subsection{Stellar parameters}
\label{ss_params}

The temperature, gravity, and surface helium abundance of the target stars were measured by fitting the observed hydrogen and helium lines
with synthetic spectra. The model spectra were computed with Lemke's
version\footnote{\small{http://a400.sternwarte.uni-erlangen.de/\~{}ai26/linfit/linfor.html}} of the Linfor program (developed originally by
Holweger, Steffen and Steenbock at Kiel University), fed with model atmospheres computed with ATLAS9 \citep{Kurucz93}. The metallicity was
set to solar, as appropriate for both recently-formed stars and hot foreground subdwarfs, whose atmospheres are enriched with heavy elements
due to diffusion processes \citep{Moehler00}. The grid of templates covered the range 8\,000$\leq T_\mathrm{eff}$(K)$\leq$35\,000,
2.5$\leq \log{(g)} \leq$ 6.0, $-$3.0$\leq \log{(\frac{N(He)}{N(H)})} \leq$ $-$1.0. The synthetic spectra  were convolved with a Gaussian
during the fitting procedure, to match the instrumental resolution of the data. The spectra of the few evolved objects hotter than 35\,000~K
were fitted using a grid of metal-free non-LTE models described in \citet{Moehler04}, calculated as in \citet{Napiwotzki97}.

The best fit and the derivation of the stellar parameters were established through the routines developed by \citet{Bergeron92} and
\citet{Saffer94}, as modified by \citet{Napiwotzki99}. The code normalizes both the model and the observed spectra simultaneously using the
same points for the continuum definition, and makes use of a $\chi^2$ test to establish the best fit. The noise in the continuum spectral
regions is used to estimate the $\sigma$ for the calculation of the $\chi^2$ statistics, which the routines use to estimate the errors of
the parameters \citep[see][]{Moehler99}. However, they thus neglect other sources of errors, such as those introduced by the normalization
procedure, the sky subtraction, and the flat-fielding. Therefore, the resulting uncertainties were multiplied by three to obtain a more
realistic estimate of the true errors (Napiwotzki 2005, priv. comm.). The lines of the Balmer series from H$_\beta$ to H12 were
simultaneously fitted, along with four He~I lines (4026~\AA, 4388~\AA, 4471~\AA, 4922~\AA). Two He~II lines (4542~\AA,
4686~\AA) were also included when visible in the spectra of hot stars. Results for all the targets are given in Table~\ref{t_data}. We could
not find a good solution for the spectrum of star SPM4 0340108821, nor could we determine if this was due to problems related to its
spectrum, or to limitations of our model grid. This object will therefore be excluded from further analysis. However, its strong and wide
He~II lines reveal that it is most likely a foreground WD ($\log{(g)}>6$, $T_\mathrm{eff}>40\,000$~K).

The lines of some stars were clearly broadened due to high projected rotational velocity. The employed routine did not include $V\sin{(i)}$
as a fit parameter, but as a fixed input quantity beforehand. We then estimated $V\sin{(i)}$ varying this value manually in such a way as to
minimize the resulting $\chi^2$ of the fit. A precise estimate of $V\sin{(i)}$ is not possible due to the relatively low resolution. In
addition, the wide Balmer lines dominated the fit statistics, but these features were intrinsically too wide for this purpose. As a
consequence, our results are reliable only for a rough distinction between stars showing no relevant evidence of rotation
($V\sin{(i)}<50$~km~s$^{-1}$), stars showing hints of rotation ($V\sin{(i)}\sim50$~km~s$^{-1}$), and fast rotators ($V\sin{(i)}\sim100$ or
200~km~s$^{-1}$). The results are given in the ninth column of Table~\ref{t_data}.

\begin{figure}
\includegraphics[width=\columnwidth,angle=-90]{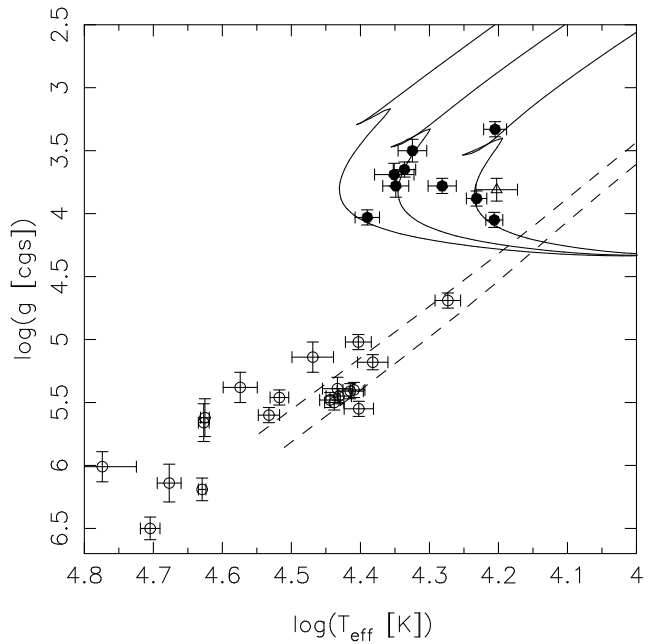}
\caption{Location of the program stars in the temperature-gravity plane. Filled and open circles represent MS and foreground compact objects,
respectively. The open triangle indicates the star whose classification is uncertain. Solid curves show solar-metallicity isochrones of 10,
20, and 50 Myr. The dashed curves indicate the zero-age and terminal-age horizontal branch.}
\label{f_tg}
\end{figure}

The surface helium abundance was assumed as a fit parameter in the procedure, and the model grid spanned a wide range from solar value
($\log{(\frac{N(He)}{N(H)})}=-1$) down to one-thousandth of solar abundance. This is appropriate for foreground subdwarfs, and this is, in
fact, a key parameter to identify them in our sample (see Sect.~\ref{ss_sdb}). However, diffusion processes are not active in MS stars, and
their surface helium abundance is in general close to solar. This parameter is therefore not very informative for these objects. On the other
hand, we found that fixing $\log{(\frac{N(He)}{N(H)})}=-1$ led to better fits of all the lines for some MS stars. We adopted the results
obtained with helium abundance fixed to solar for these objects, and this parameter has no associated error in Table~\ref{t_data}. We
nevertheless verified that our final results were not affected by this choice as, for example, distance moduli and ages changed by less than
1--1.5$\sigma$.


\section{Results}
\label{s_res}

\subsection{MS stars and foreground objects}
\label{ss_sdb}

The RV distribution of target stars is clearly bimodal, with a huge gap between ten high-velocity stars (RV$>160$~km~s$^{-1}$) and the rest
of the sample (RV$<55$~km~s$^{-1}$). This dichotomy suggests an easy separation between LMC members and foreground stars, because the RV of
the LMC \citep[262.1$\pm$3.4~km~s$^{-1}$][]{vanderMarel02} is very high compared to the typical RV of Galactic stars along the same line of
sight. However, our single-epoch measurements alone are not enough to uniquely identify LMC stars from Galactic interlopers, since the
incidence of close binaries is very high among both massive OB stars \citep{Sana12} and subdwarfs \citep{Maxted01}, although the binary
fraction for old halo sdB's could be much lower than in the disk \citep{Moni08,Han08}. Galactic runaway stars, on the other hand, are
predominantly single objects \citep{Mason98}, but they can show anomalously high RV due to their intrinsic nature of being high-velocity
stars. However, the measured stellar parameters provide a straightforward identification of sdB's and WDs. In fact, their surface gravity
is higher than that of MS stars at the same temperature for $T_\mathrm{eff}>15\,000$~K \citep[see, e.g.,][]{Salgado13}, and their loci in
temperature-gravity space are well separated. The atmospheres of sdB's with $T_\mathrm{eff}<35\,000$~K are also depleted of helium
\citep[typically $\log{(\frac{N(He)}{N(H)})}<-1.5$,][]{Moni09}, at variance with the close-to-solar values expected for MS stars. In
addition, while fast rotators are common in the MS, they are extremely rare among sdB's \citep{Geier12}.

\begin{figure}
\includegraphics[width=\columnwidth,angle=-90]{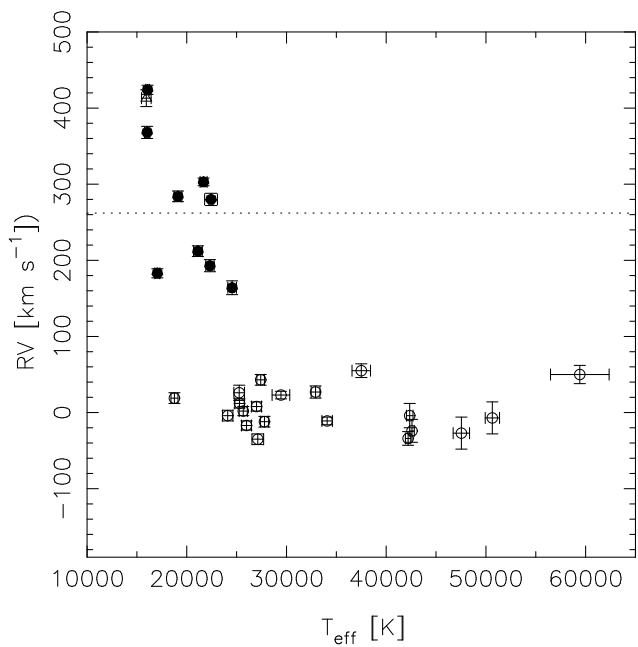}
\caption{RV of target stars as a function of their effective temperature. The symbols are as in Fig.~\ref{f_tg}. The dotted line indicates
the systemic RV of the LMC.}
\label{f_trv}
\end{figure}

Figure~\ref{f_tg} shows the position of target stars in the temperature-gravity plane. Solar-metallicity isochrones from \citet{Bressan12}
are also shown as solid curves. Dashed lines indicate the zero-age and terminal-age horizontal branch, which enclose the loci where stars
spend 90\% of their helium core-burning stage. Twenty stars lie on the loci of sdB's and WDs, and their surface gravity is too high for MS
objects ($\log{(g)}>4.7$). All but two of the high-gravity stars with $T_\mathrm{eff}<40\,000$~K are strongly helium-depleted
$\log{(\frac{N(He)}{N(H)})}<-2$, while surface helium abundance increases for the hottest stars up to super-solar values, as is usual for
O-type subdwarfs and hot WDs \citep[e.g.,][]{Edelmann03}. Only two of these stars show hints of rotation (SPM4 0870012815 and
SPM4 0570014281). All these objects are therefore classified as foreground stars. The gravity of the coolest star of our sample
(SPM4 0140123632) is too low to be an sdB. However, its very low helium abundance is incompatible with it being an MS star. While its
classification should be regarded as dubious, we interpret it as a post-horizontal branch star, evolving at lower gravities toward the
Asymptotic Giant Branch after exhaustion of helium in the core. It should be a foreground object even in this case, because its absolute
magnitude would have to be too bright for a post-HB star at its temperature to place it at the distance of the LMC
\citep[$(m-M)_0\approx18.5$,][]{Freedman01}. The remaining nine targets are clearly MS stars, as deduced from their position in the
temperature-gravity plane and their close-to-solar surface helium abundance. Six of them also show evidence of non-negligible rotation. The
classification of each target (``sdB/WD'', ``post-HB?'', or ``MS'') is given in in the last column of Table~\ref{t_data}.

The RVs of our targets are shown in Fig.~\ref{f_trv} as a function of effective temperature, where we indicate foreground compact objects
and MS stars with empty and full symbols, respectively. The Figure reveals that the RV of all foreground sdB's/WDs is lower than
55~km~s$^{-1}$, while the aforementioned high-velocity stars are all MS stars, with the addition of the uncertain post-HB object. The
foreground stars will not be considered further, and we will focus on the nine MS stars hereafter.

\subsection{Distances and ages}
\label{ss_distage}

\begin{table*}
\centering
\caption{Reddening, age, distance modulus, V$_{lsr}$ and Magellanic coordinates of the target MS stars.}
\label{t_aged}
\begin{tabular}{c c c l c c r r }
\hline
\hline
ID & SPM ID & E($B-V$) & age & $(m-M)_0$ & V$_{lsr}$ & $\xi_{\Lambda_M}$ & $\eta_{B_M}$ \\
 & & mag & Myr & mag & km~s$^{-1}$ & $\degr$ & $\degr$ \\
\hline
005 & 0030025059 & 0.10 & 50$\pm$20 & 15.52$\pm$0.27 & 174 & $-$2.281 & $-$12.611 \\
292 & 0320115607 & 0.10 & 11$\pm$6  & 18.36$\pm$0.18 & 153 & $-$4.247 & $-$2.510  \\
290 & 0320018938 & 0.10 & 20$\pm$5  & 17.62$\pm$0.26 & 182 & $-$4.240 & $-$2.519  \\
403 & 1250015381 & 0.01 & 20$\pm$5  & 18.62$\pm$0.38 & 198 & $-$7.065 &   12.871  \\
405 & 1280025746 & 0.04 & 20$\pm$5  & 18.45$\pm$0.24 & 288 &  3.910   &   13.412  \\
406 & 1280031295 & 0.05 & 45$\pm$10 & 18.60$\pm$0.27 & 353 &  4.220   &   14.491  \\
487 & 3380011095 & 0.07 & 55$\pm$30 & 15.48$\pm$0.29 & 407 &  9.457   &   33.541  \\
390 & 0920126094 & 0.11 & 35$\pm$8  & 18.50$\pm$0.21 & 271 &  9.310   &    4.555  \\
307 & 0350232289 & 0.16 & 18$\pm$5  & 18.79$\pm$0.29 & 268 &  7.447   & $-$1.609  \\
\hline
\end{tabular}
\end{table*}

\begin{figure}
\includegraphics[width=\columnwidth,angle=-90]{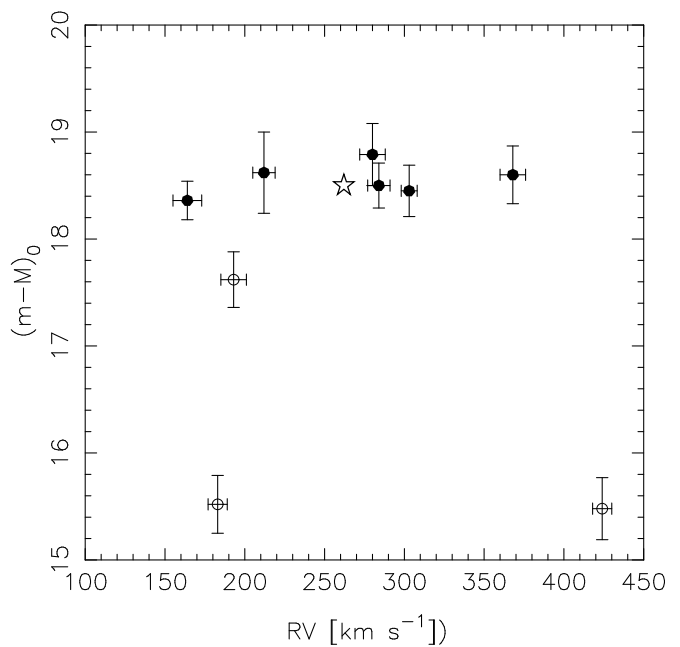}
\caption{Distance modulus of the target MS stars as a function of their RV. Filled and open circles indicate the LMC members and the
foreground runaway stars, respectively. The star symbol indicates the value of the LMC.}
\label{f_drv}
\end{figure}

The age of the MS stars was estimated comparing their position in the temperature-gravity plane with solar-metallicity isochrones, as shown
in Fig.~\ref{f_tg}. Varying the stellar parameters within their error bars caused small variations in the derived age, of the order of 10\%.
A larger uncertainty was introduced by the assumption of the isochrone metallicity, as a change of $\pm$0.5~dex affected the results by
$\mp$20\% or more. We therefore assumed this uncertainty as the final error.

The observed $V$ magnitudes were de-reddened by means of the line-of-sight values given by the \citet{Schlegel98} maps, as corrected by
\citet{Bonifacio00}, assuming a standard reddening law ($R=\frac{A_V}{E(B-V)}=3.1$). An estimate of the distance modulus was thus obtained,
by comparison with the absolute magnitude in the $V$ band obtained from the same solar-metallicity isochrones. The assumption of
$[\mathrm{Fe/H}]=\pm0.5$ for the theoretical isochrones shifted the absolute magnitude by $\mp$0.2~mag at most ($\mp$0.12~mag on average).
A larger but comparable uncertainty on $(m-M)_0$ was induced by the observational errors of the stellar parameters (0.23~mag on average),
and the two values were quadratically summed to obtain the final errors. The results for reddening, age, distance modulus, and line-of-sight
velocity with respect to the Local Standard of Rest V$_{lsr}$ are given in Table~\ref{t_aged}, where we also add an internal ID (first
column) and sky coordinates (see Section~\ref{s_ana}).

Figure~\ref{f_drv} shows the RV of our MS objects as a function of their distance modulus, compared to the LMC values
RV$_\mathrm{LMC}$=262.1$\pm$3.4~km~s$^{-1}$ \citep{vanderMarel02} and $(m-M)_\mathrm{0,LMC}$=18.5 \citep[distance $d$=50~kpc;][]{Freedman01}.
Two stars are clearly foreground objects ($(m-M)_0\approx$15.5, d$\approx$12.5~kpc). A third more distant star ($d\approx$33~kpc), whose
$(m-M)_0$ is incompatible with the LMC at the 3.3$\sigma$ level, is also likely not a LMC member. These three stars are probably Galactic
runaway stars, as also suggested by their high recession velocity. The distance modulus of the six remaining stars agrees with the LMC value
within 1$\sigma$. Therefore, we consider these six young stars as being members of the LMC.


\section{Analysis}
\label{s_ana}

In what follows, we adopt the Magellanic coordinate system introduced by \citet{Nidever10}, in which $\Lambda_{M}$ is the longitude along the
Magellanic Stream, positive toward the direction of motion of the LMC (i.e., toward the Galactic plane), and $B_M$ is the Magellanic latitude.
Our adopted center for the LMC is (RA, Dec)=($81\fdg90, -69\fdg86$), from \citet{vanderMarel02}. This center, however, does not correspond to
the origin of the Magellanic coordinate system, but is slightly offset: ($\Lambda_{M,LMC}$, $B_{M,LMC}$) = ($0\fdg215 ,2\fdg330$). We adopt
this system since it reflects the orbital motion of the LMC, and --- at the location of the LMC --- it is closely aligned with the equatorial
system (a $\sim2\degr$ offset). Finally, the spherical ($\Lambda_{M}$, $B_{M}$) coordinates are projected onto a plane tangent to the sky at
($\Lambda_{M}$, $B_{M}$)=($0\degr, 0\degr)$, a gnomonic projection with coordinates ($\xi_{\Lambda_M}, \eta_{B_M}$). Units for sky coordinates
are degrees throughout the paper.

\subsection{Matching with Existing Catalogs of Clusters and Associations}
\label{ss_match}

\begin{figure*}
\includegraphics[width=7cm,angle=-90]{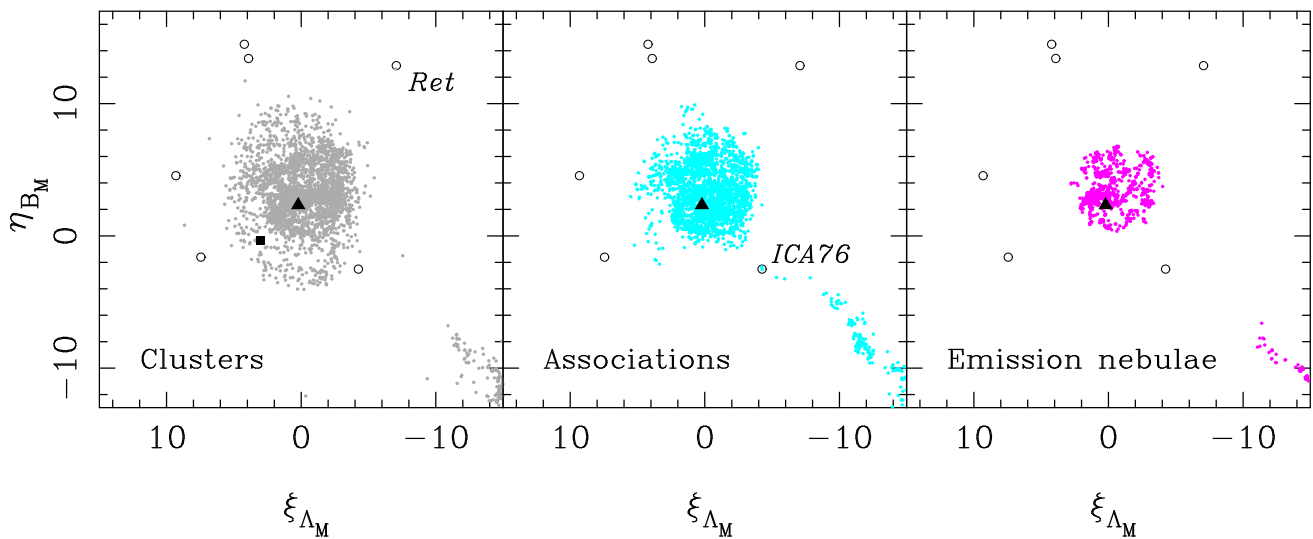}
\caption{Spatial distribution of our six young stars (open circles) compared to that of clusters (left), young associations (middle),  and
emission nebulae (right) from \citet{Bica08}. The center of the LMC is marked with a triangle. The positions of the globular cluster Reticulum
and the stellar association ICA76 are also labeled. The dark square marks the newly-discovered, loose cluster by \citet{Piatti16}.}
\label{f_bica}
\end{figure*}

We have cross-matched our six stars with the catalogs of \citet{Bica08}. In Fig.~\ref{f_bica} we show the spatial distribution of our six
stars, and of the catalog of clusters (left panel), young associations (middle panel) and emission nebulae  (right panel). Target $\#403$ is
$15\arcmin$ from the center of the globular cluster Reticulum. However, this star is not a cluster member, since it is a young star, and its
RV of $212\pm7$~km~s$^{-1}$ is different from that of the cluster \citep[$241.5\pm1.5$~km~s$^{-1}$][]{suntzeff92}. In Fig.~\ref{f_bica}, we
also mark the newly-found, loose cluster by \citet{Piatti16}. This $280$~Myr open cluster is at a distance some 11.3~kpc larger than that of
the LMC. However, its sky location is well within the distribution of the clusters catalogued by \citet{Bica08}.

The target $\#292$ is within the stellar association ICA76 (aka IDK6) as given in \citet{Bica08}, \citet{Irwin90}, and \citet{Batt92}.
Specifically, $\#292$ is at $2\arcmin$ from the center of ICA76, which has a radius of $\sim7\arcmin$ \citep{Bica08}. We note that star
$\#290$ is also located within the projected radius of ICA76, but the distance modulus determined here (Tab.\ref{t_aged}) places it in the
foreground. \citet{Dem98} determine a distance modulus of $18.63\pm0.10$ to ICA76, and an age of 10-25 Myr. This agrees reasonably well with
the distance and age of star $\#292$ (see Tab.\ref{t_aged}). \citet{Dem98} also quote previous RV measurements of these two stars: they are
$194$~km~s$^{-1}$ and $270$~km~s$^{-1}$ for $\#290$ and $\#292$, respectively. Ther RV for $\#290$ agrees with our measurement
($193\pm8$~km~s$^{-1}$), however that for $\#292$ does not ($164\pm4$~km~s$^{-1}$). While we don't have an explanation for this discrepancy,
one reason may be binarity.

We have cross-matched our six stars with the recent DES cluster catalog by \citet{Pieres15}, and found no matches. We have also inspected
$15\arcmin\times15\arcmin$ GALEX and Digital Sky Survey B images centered on each of our six stars. Excepting $\#292$, where the stellar
association ICA76 is clearly visible, the remaining five stars appear isolated. To conclude, our six young stars are much further out in
projected distance from the LMC's center than the catalogued clusters, associations, and emission nebulae. Only one star in our sample is a
likely member of a known young stellar association located in the Bridge, while the other five objects cannot be identified as being
associated with any known stellar aggregate.

\subsection{Spatial Distribution}
\label{ss_spdist}

\begin{figure}
\includegraphics[width=12cm,angle=-90]{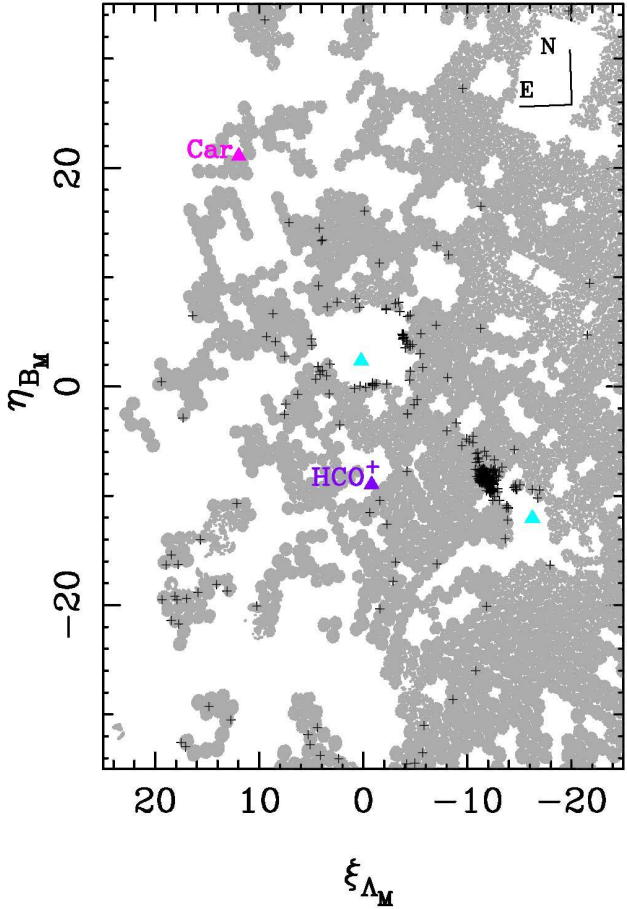}
\caption{Area coverage of the \citetalias{Dinescu12} study to select candidates (grey background). The round footprint of GALEX is easily
visible; the gaps in area coverage are mainly due to the GALEX GR5 coverage. Young, OB-type star candidates from \citetalias{Dinescu12} are
shown with crosses. The centers of the LMC and SMC are shown with light-blue triangles. Two notable sky locations are also indicated and
labeled: the Carina dwarf galaxy and the first detection of HCO$^+$ absorption in the Magellanic System by \citet{murray15}. North and East
directions are also indicated.}
\label{f_map_sample}
\end{figure}

In Figure~\ref{f_map_sample} we show the area coverage of the \citetalias{Dinescu12} study to select OB-type candidates in the region of the
Clouds. Gaps in the coverage are primarily due to the GALEX GR5 survey. Candidates from \citetalias{Dinescu12} are shown with crosses: clearly,
only the outer regions of the Clouds were probed by \citetalias{Dinescu12}. The Galactic plane is toward the left of this Figure. We also mark
two useful locations in this plot: the center of Carina dwarf galaxy, and the first detection of HCO$^+$ absorption at the leading edge of the
Magellanic Bridge by \citet{murray15}. \citet{Munoz06} found giant stars, likely members of the LMC in the Carina foreground, thus hinting at
the large spatial extent of the Cloud, while the detection of HCO$^+$ is indicative of star formation. Figure~\ref{f_map_sample} is meant to
caution the reader as to the areal incompleteness.

In Fig.~\ref{f_map_nh} we show the map of the H~I column density in our region centered on the LMC. The H~I data are from the GASS~III cleaned
data described in \citet{Kalberla15}, within the velocity range V$_{lsr}$=100 to 450~km~s$^{-1}$. OB-type candidates from \citetalias{Dinescu12}
are shown with grey crosses. Stars observed spectroscopically in this study are shown with circles, young stars being highlighted in green, and
the six LMC members are marked with a square. As in Fig.~\ref{f_map_sample}, Carina's center and the location of  HCO$^+$ detection are
indicated. The thick blue line marks the newly found 10-kpc tidal feature/arm by \citet{mackey16} in the DES. The thick blue dotted line marks
the western extension of this arm, a feature less conspicuous than the arm.

Our six stars are surrounding the LMC out to $\sim13\degr$ from its center. Star $\#292$ which belongs to association ICA76 is clearly located
in the Bridge, and it is the closest in projected distance to LMC's center. Two stars, $\#405$ and $\#406$, are very close to each other (within
$\sim1\degr$), while the remaining three are at various position angles. Except for $\#292$ which is in the Bridge, the stars are in regions of
low H~I column density, $N_{HI}<10^{19} $ cm~$^{-2}$. It is intriguing that the two close together stars, $\#405$ and $\#406$, are only some
$5\degr$ behind the stellar tidal arm found by \citet{mackey16}. \citet{mackey16} also indicate the existence of a western extension of this
tidal arm in their Fig.~2 (the blue dotted line in our Fig.~\ref{f_map_nh}). This extension, if real, is very near to our star $\#403$, with the
star slightly closer to the LMC center than the extension. We also note that \citet{Besla16} found stellar arcs and spiral arms in the northern
periphery of the LMC. These features are within $\sim8\degr$ of the LMC's center, specifically in region C of \citet{Besla16}, their Fig.~4. Our
study does not properly sample these structures, since we focused on more distant objects (see  Fig.~\ref{f_map_nh}). None of our candidates
lying close to the HCO$^+$ detection proved to be LMC members, but that region of the sky is affected by a large areal gap in the survey data
(see Fig.~\ref{f_map_sample}).

\subsection{Testing a Ring-like Configuration}
\label{ss_ring}
 
\begin{figure}
\includegraphics[width=11cm,angle=-90]{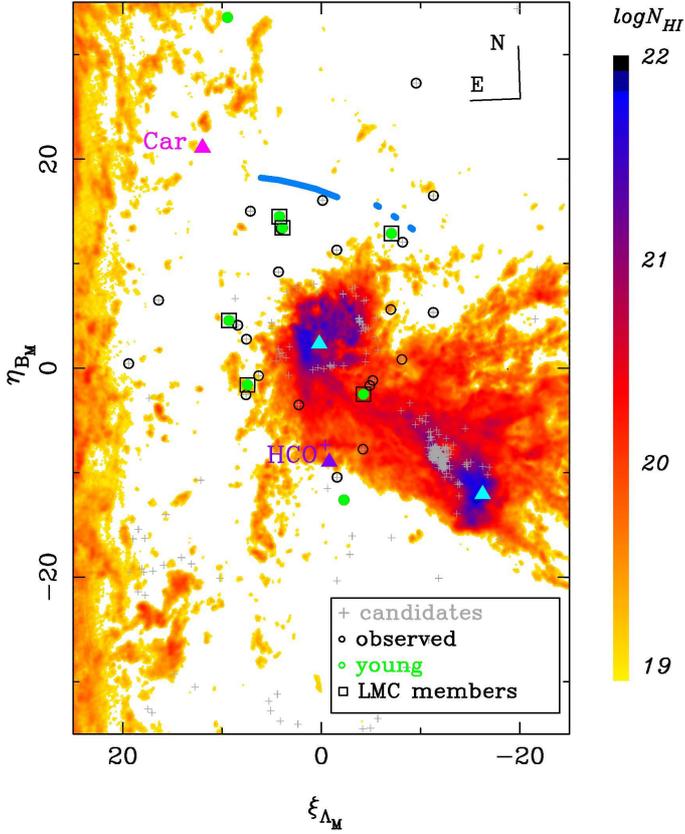}
\caption{Map of the H~I column density from the GASS survey within V$_{lsr}$=100 to 450~km~s$^{-1}$ in the region of interest. The centers of
the Clouds are shown with light blue triangles. OB-type candidates from \citetalias{Dinescu12} are shown with grey crosses. Circles represent
the stars observed spectroscopically in this study. Filled green circles indicate the young stars. Squares indicate the six young stars that
are members of the LMC. The thick blue curve indicates the newly found 10-kpc tidal feature/arm by \citet{mackey16} in the Dark Energy Survey.
The thick dotted blue curve shows the western extension of this arm.}
\label{f_map_nh}
\end{figure}

Our five stars (excluding the ICA76 member) appear to be in a half-ring configuration, with the ring's center some $3\degr$ northward of LMC's
center, and spanning some $\sim140\degr$ in position angle (see Fig.~\ref{f_map_nh}). Note that some 10-12 candidates at moderate
$\sim7\degr-10\degr$ distance from the LMC's center are foreground stars (see open circles in Fig.~\ref{f_map_nh}).  This leaves a gap between
a region of inner star formation, as traced by emission nebulae and young associations (Fig.~\ref{f_bica}), and the off-center, ring-like
configuration of our five stars.  Also, the five stars coincide with the ``near'' part of the LMC disk (see Fig.~\ref{f_vel_diff}). Alas, our
distance uncertainties are inadequate to explore the ``nearness'' of these five stars.

Our targets were selected from the \citetalias{Dinescu12} candidates located between 6$\degr$ and 33$\degr$ from the center. Hence, the
spatial distribution of the observed sample has already some degree of circular symmetry about the LMC center. One might wonder to what extent
the ringlike feature traced by the five young LMC stars could be an effect induced by the underlying distribution of the selected targets. That
is, while the stars found to be young LMC members appear to lie in the form of a ring, are they really that much more ``ringlike'' than other,
random groupings of five stars that might have been drawn from the observed sample? To answer this question, we perform a Monte Carlo test to
quantify the statistical significance of the observed configuration.

From the 31 spectroscopically observed stars we discard four that are located in the Bridge. From this set of 27 stars, we randomly
select five stars and project their sky positions into the LMC disk plane. The geometry of the disk is adopted from \citet{olsen11}, with an
inclination angle of $34\fdg7$, and a position angle of the line of nodes of $142\fdg0$ (see also Sec.~\ref{ss_veldist}). We search in a grid
around the LMC's center for a ring-center position that gives the five stars a minimum in $\delta r/r_{ave}$, where $r_{ave}$ is the average
radius of a circle they define, and $\delta r$ is the root-mean-square scatter of the stars' positions from a perfect circle. The optimal
center and $r_{ave}$ define the ``best-fit'' circular ring for this group of five stars. We also calculate the arc length that the five stars
subtend along the best-fit circle. The search for the ring center is done in steps of $0\fdg25$ within $7\degr$ of the LMC's center. We perform
1000 such random, five-star trials and store the resulting values of $\delta r/r_{ave}$ and arc length. Then, we repeat this 1000-trial run a
total of ten times. For each run, we determine how many groups of five stars form a ``better'' circle than the actual five LMC stars, i.e.,
have $\delta r/r_{ave}$ smaller than that of the five LMC stars. We find that the angular arc length of our five stars is fairly typical of
randomly drawn ones. However, on average, only $1.35\%$ of randomly drawn groups form an arc of a circle with $\delta r/r_{ave}$ lower than
that of the five LMC stars. This result suggests that an accidental hypothesis can be ruled out at the 98.65\% confidence level. The ringlike
configuration of these five stars is therefore unlikely to be due to chance, and a physical basis is suggested.
We speculate as to the possible origin of such a structure in Sec.~\ref{ss_origin}.

The ring parameters found for the five LMC stars are: average radius  $r_{ave}=13\fdg1$, distance from the LMC's center $\rho=3\fdg0$, and
position angle arclength $\Delta PA=145\degr$. At a distance of 50~kpc for the LMC, the average radius is 12~kpc, and the distance from LMC's
center is 2.6$\pm$0.3~kpc. All of these parameters are in the LMC's disk plane.

While our statistical test shows it is unlikely that the ring is mere coincidence -- and we proceed to derive the geometry in the plane of the
LMC as if this were a real structure -- this is all based on five stars situated over a rather large portion of sky. Determining whether these
stars actually reside in the LMC disk will require additional observations (primarily astrometric) that are not available at this time.

\subsection{Velocity Distribution}
\label{ss_veldist}

\begin{figure*}
\includegraphics[width=10cm,angle=-90]{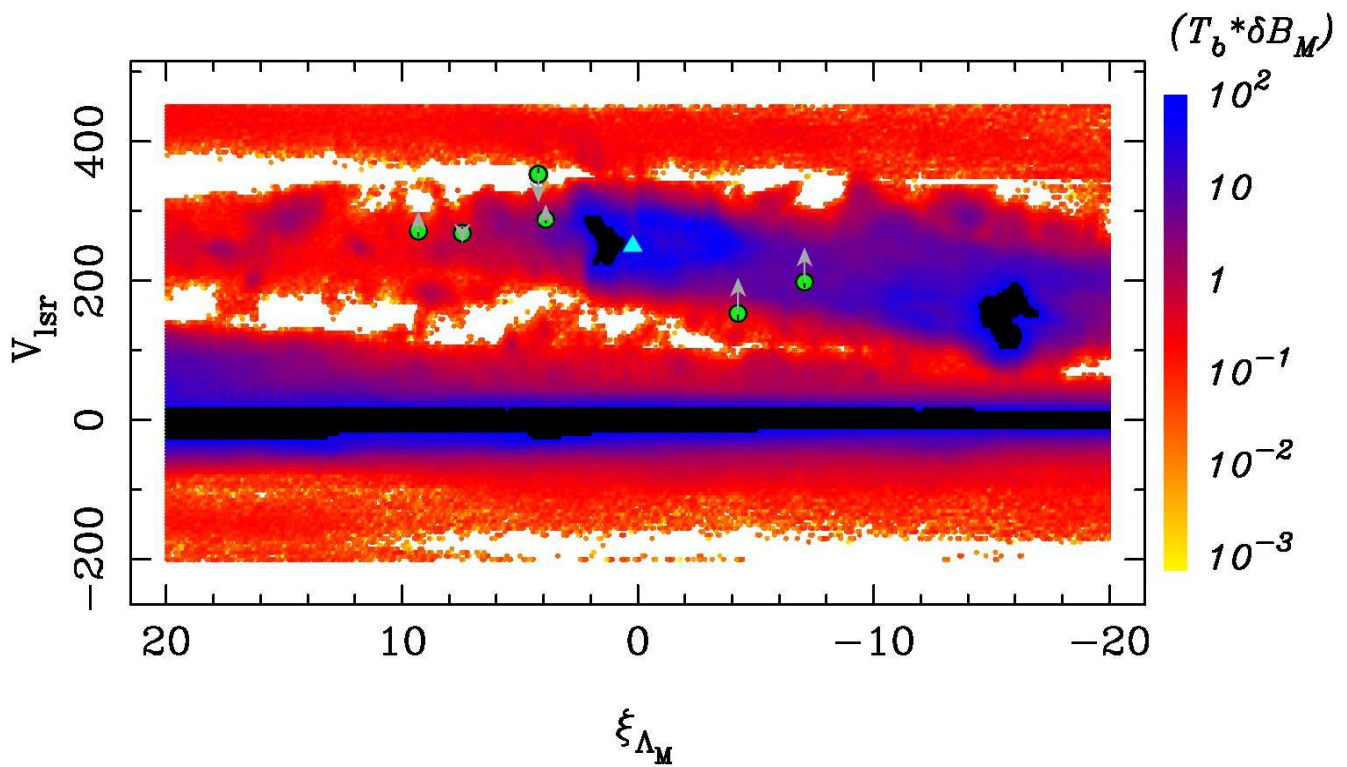}
\caption{V$_{lsr}$ as a function of $\xi_{\Lambda_M}$ along the LMC's orbit. The color scale shows the temperature brightness integrated
along $\eta_{B_M}$ for the range -30$\degr$ to 30$\degr$. Black regions indicate values larger than $10^2$ K deg, while white regions
indicate values less than  $10^{-3}$ K deg. Green symbols show our young LMC members. The LMC's center is represented with a triangle.
The arrows indicate the difference in velocity between our measurements and the predictions of a model of the LMC disk rotation, at each
star's position.}
\label{f_vel_gas}
\end{figure*}

We compare our measured radial velocities for the LMC members with that of the H~I gas \citep{Kalberla15} in Fig.~\ref{f_vel_gas}. Plotted is
the local standard of rest radial velocity V$_{lsr}$ as a function of $\xi_{\Lambda_M}$, the coordinate along the orbit of the LMC. The color
scale represents the temperature brightness integrated along $\eta_{B_M}$ for the range $-30\degr$ to 30$\degr$. The location of the Galactic
disk, and of the central parts of the Clouds are clearly visible. Our young LMC members are shown with green symbols. They follow the overall
pattern of the gas motions, with the two most deviant points being star $\#292$ at $\xi_{\Lambda_M}=-4\fdg247$, and star $\#406$ at
$\xi_{\Lambda_M}=4\fdg220$. Star $\#292$ is a member of ICA76, located in the Bridge; and in Section~\ref{ss_match} we noted a previous
velocity measurement for this star (RV=270~km~s$^{-1}$, or V$_{lsr}$=259~km~s$^{-1}$), which would bring it in better agreement with the gas
motion.

We also compare our velocity measurements with the predictions of a disk model for the LMC. We adopt the geometry and rotation curve of the
disk from \citet{olsen11}; these were determined from the analysis of red supergiants, a relatively young population. The adopted distance to
the LMC is 50 kpc, and its RV is 263 km~s$^{-1}$. The inclination angle of the disk is $34\fdg7$, and the position angle of the line of nodes
is $142\fdg0$. The model assumes circular motions only, and we use the formalism from \citet{vanderMarel02} to calculate velocities at a given
position angle and angular distance from the LMC's center. The arrows in Fig.~\ref{f_vel_gas} point to the velocity model predictions at each
star's location in ($\xi_{\Lambda_M}$, $\eta_{B_M}$), i.e., the length of the arrow is the velocity difference between observation and model.
These differences are rather modest, indicating that the velocities of the stars are not too deviant from simple, circular motion in the disk
of the LMC.

\begin{figure}
\includegraphics[width=\columnwidth,angle=-90]{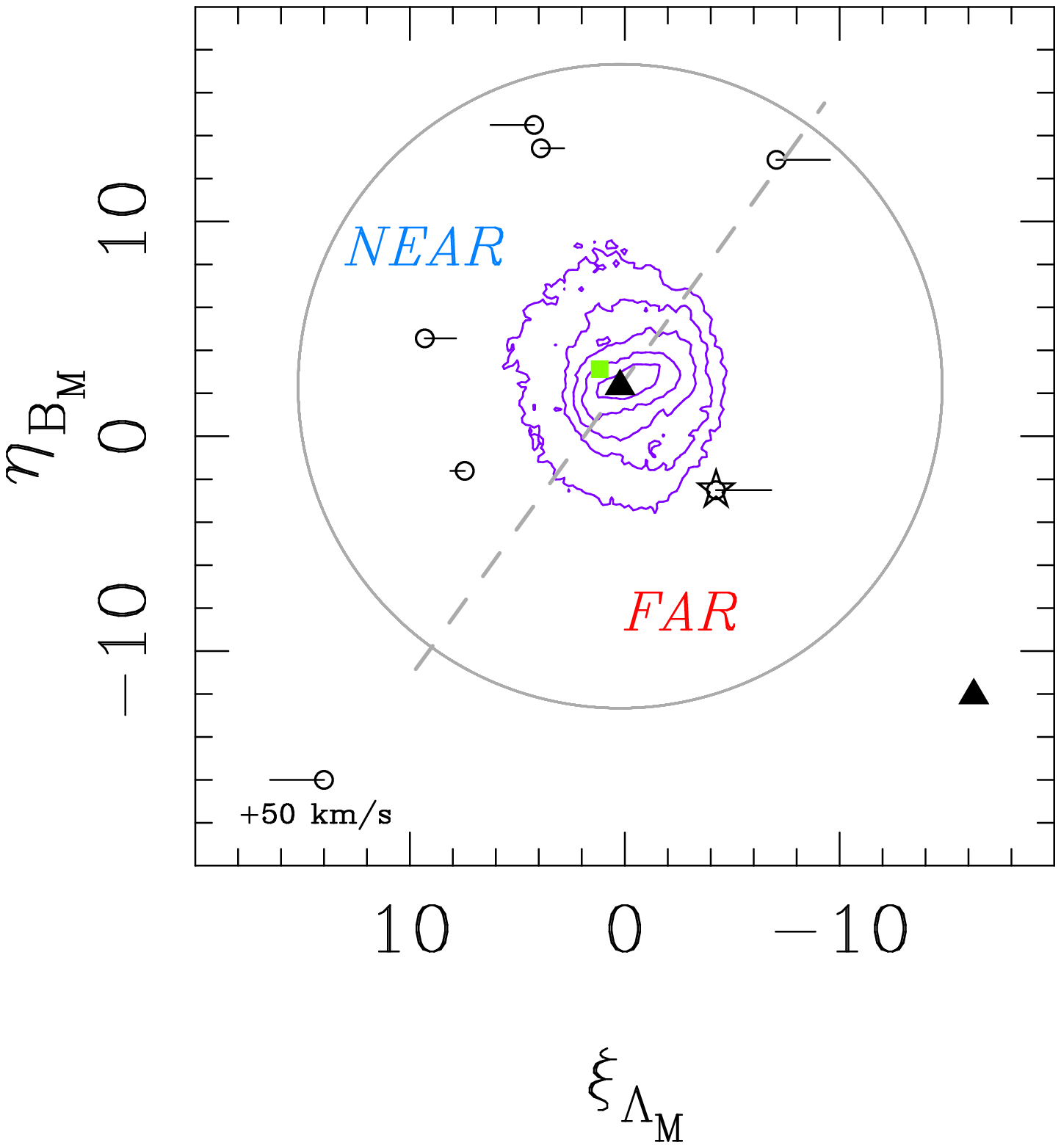}
\caption{The sky distribution of the six young LMC members with V$_{lsr}$ differences (observed - model) shown as horizontal vectors. A
vector of +50 km~s$^{-1}$ is shown in the bottom left corner for scale. Triangles mark the centers of the LMC and SMC. The star symbol
shows the stellar member of ICA76. The green square indicates the location of 30 Doradus. Contours represent isodensity curves of M giants
in the LMC selected from 2MASS. The line of nodes of the disk is shown with a dashed line, with the near and far sides of the disk labeled.
The circle highlights a radius of $15\degr$ from LMC's center.}
\label{f_vel_diff}
\end{figure}

To explore whether the velocity differences have any correlation with distance from LMC center, or with the position angle, we plot these
velocity differences at the sky location of each star in Fig.~\ref{f_vel_diff}. For reference, we show isodensity contours of M giants in
the LMC selected from 2MASS as in \citetalias{Dinescu12}. The LMC's bar is easily visible in the inner regions. We also mark the location
of the star-forming region 30 Doradus with a green square. The line of nodes of the disk is shown with a dashed line, and the ``near''
and ``far'' halves of the disk are labeled. The six young LMC members are shown with circles, and the member of ICA76 is highlighted with
a star symbol. A circle with a radius of $15\degr$ centered on the LMC is also shown. Velocity differences do not show a correlation with
either projected distance from LMC center or with position angle. The average of these differences is $-16\pm15$~km~s$^{-1}$, with a
standard deviation of 36 km~s$^{-1}$. If we exclude star $\#292$, the member of ICA76, we obtain an average velocity difference of
$-9\pm16$~km~s$^{-1}$, with a similar standard deviation of 36 km~s$^{-1}$. 

\subsection{Origin of the Young Stars}
\label{ss_origin}

Inspecting Table~\ref{t_aged}, we see that star $\#292$ is 11~Myr old, the youngest in our sample of six stars. The ages of the remaining
five stars range between 18 and 45~Myr, with an average of $\sim30$~Myr. If we assume the standard deviation of the line-of-sight
velocities is representative of the deviations from simple LMC disk rotation for the tangential components as well, we can estimate the
maximum distance in the sky these stars will travel in 30~Myr. For 36~km~s$^{-1}$ (Section~\ref{ss_veldist}), this distance amounts to
1~kpc, or $\sim1\degr$ at the distance of the LMC. Thus, in absence of precise proper motions for these stars, we argue that line-of-sight
velocities of the young LMC members indicate that they have not traveled far during their lifetimes, and their current locations are
representative of their birthplace. This points to {\it in situ} formation rather than being runaways from star-forming regions and stellar
associations in the inner LMC. Moreover, stars $\#405$ and $\#406$ being so close to one another further argues against the likelihood of a
runaway mechanism, since one would assume random directions for the ejecting stars.

Our findings are rather unexpected, since they point to star formation in the far outskirts of the LMC where gas density is currently low,
$N_{HI}<10^{19}$~cm~$^{-2}$ (Fig.~\ref{f_map_nh}). Figure~\ref{f_bica} clearly shows this as well; notice the small, inner LMC region
occupied by emission nebulae and young stellar associations compared to the distribution of our stars. That the gaseous disk of the LMC is
truncated when compared to the intermediate-age/old stellar disk is no longer a novelty since the work of \citet{Saha10}. Various
theoretical models \citep[see][and references therein]{Hammer15} attribute this morphology primarily to truncation caused by ram pressure
exerted by the Galactic hot halo gas.  Traditional star-formation studies of the LMC disk, based on color-magnitude diagrams, indicate that
the young star-forming epoch continues to the present time only in the inner $\sim5\degr$ of the disk \citep{Meschin14}.

Whence these $\sim20-50$~Myr stars in the far outskirts? One intriguing clue as to a possible origin is their half-ring spatial distribution,
which was shown in Sec. 4.3 to be unlikely a mere random configuration. If these are, in fact, in the form of a circular ring in the plane of
the LMC disk, it suggests an interesting possible formation scenario.

N-body/SPH simulations of collisions between a barred disk galaxy and a small (20\% mass) satellite \citep{Beren03} show that an off-center
collision produces expanding ring structures, off-center bars, and other asymmetries in the stars and gas. Typically, the impact produces a
density wave in the host's disk, which first becomes apparent in the gas. If the impact is off center, the symmetry of the gaseous ring is
broken when it encounters spiral arms. In our case, the SMC had a recent collision with the LMC's disk. The particulars of this collision
--- impact parameter and angle --- will dictate the morphology of the outskirts of the LMC's disk
\citep[see e.g.,][and references therein]{Besla16}.  It is thus conceivable that the formation of our five young stars which appear in a
ring-like configuration was triggered by the recent collision between the SMC and the LMC's disk. At the time the \citet{mackey16} study was
completed, the DES did not cover the eastern portion of the outskirts of the LMC where our stars $\#307$ and $\#390$ lie; it would be very
informative to have that area covered by deep, contiguous photometric surveys.

The impact parameter of the collision between the SMC and the LMC disk appears to be moderate to small. Based on a simple orbit integration
\citetalias{Dinescu12} find an impact parameter of $\sim12$~kpc as an upper limit, and an impact angle of $\sim35\degr$ (where $0\degr$ is
coplanar). From dynamical models of the LMC-SMC interaction, \citet{Besla16} find the impact parameter to be $< 10$ kpc in order to reproduce
the arcs found in their study, while \citet{Besla12} find it to be $2-5$~kpc in order to warp the bar out of the LMC disk plane. The center
of our half-ring structure can be conjectured as the location of the impact between the SMC and LMC's disk \citep[see e.g.,][]{Mapelli12}.
In Section~\ref{ss_ring} we determine the physical distance between the LMC's center and that of the five-star ring to be $2.6\pm0.3$~kpc, a
value that is toward the low end of the estimated impact parameter from models of the interaction. The uncertainty in this value is derived
from $\delta r$ (see Section~\ref{ss_ring}), i.e., it is based on the fit of the five LMC stars. Nevertheless, our observations bring to
light an as yet unknown aspect of the LMC-SMC interaction: the formation of young stars in a half-ring configuration in the outskirts of
LMC's disk. These observations should be useful in constraining the geometry of the collision from a totally new viewpoint, as well as
provoking models of star formation in low gas environments.


\section{Summary}
\label{s_concl}

We have measured radial velocites and stellar parameters for thirty-one stars from the \citetalias{Dinescu12} sample, candidates for being
young OB stars in the periphery of the LMC; specifically between $6\degr$ and $33\degr$ from the LMC's center. Twenty-one of our targets
are foreground subdwarfs or white dwarfs, and one is an object of uncertain classification, likely a post-HB star. The remaining nine targets
are young, hot, MS stars, whose distances and ages are determined in our study. We find that three of them are foreground to the LMC,
probably runaway stars from the Galactic disk, while six of our targets are recently-formed stars at the distance of the LMC. The youngest of
these (11~Myr) is a member of a known stellar association, ICA76, located in the Bridge. The remaining five stars span a narrow age range,
between 18 and 45~Myr. These five MS objects are located between $\sim8\degr$ and $13\degr$ from LMC's center in regions of low H~I density,
and appear to have formed in isolation. Their radial velocities follow the H~I gas motions in the Magellanic system, and are not too deviant
from the predictions of an LMC disk model. Specifically, velocity differences with the model are on average 36~km~s$^{-1}$, which imply a
travel distance of at most 1~kpc ($\sim1\degr$ at the LMC's distance) during their lifetime. We argue that these stars were formed {\it in
situ} rather than being runaways from the inner disk of the LMC. Their spatial configuration appears ring-like, and is suggestive of a
formation process triggered by the recent collision of the SMC with the LMC disk at a modest, $4$-kpc impact parameter. Highly accurate
trigonometric parallaxes and absolute proper motions for these young stars, as is expected within the next few years from the Gaia mission,
should readily confirm or refute this possible explanation. Regardless of their apparent geometric configuration, these newly identified LMC
members represent a feature that, unlike previously discovered structures in the periphery of the disk, is traced by young stars.


\section*{Acknowledgements}
This study was based on observations gathered at the 6.5 meter Magellan Telescopes located at Las Campanas Observatory, Chile (program ID
CHILE-2015A-029). C.M.B. acknowledges support from FONDECYT through regular project 1150060. R.A.M. acknowledges support from the Chilean
Centro de Excelencia en Astrof\'isica y Tecnolog\'ias Afines (CATA) BASAL PFB/06 and from the Project IC120009 Millennium Institute of
Astrophysics (MAS) of the Iniciativa Cient\'ifica Milenio del Ministerio de Econom\'ia, Fomento y Turismo de Chile. L.Z. acknowledges
supports from the Chinese Academy of Sciences (CAS) through a CAS-CONICYT Postdoctoral Fellowship and partial supports from  NSFC grants
11303037 and 11390371. V.K. acknowledges the financial support by the RFBR grant 15-02-06204-a.

\bibliographystyle{mnras}
\bibliography{LMC_Halo}

\bsp	
\label{lastpage}
\end{document}